\documentclass[%
preprint,
showpacs,preprintnumbers,
 amsmath,amssymb,
 aps,
prd,
]{revtex4-1}

\usepackage{graphicx}
\usepackage{dcolumn}
\usepackage{bm}

\begin{document}

\preprint{}

\title{Thermal Fluctuations Effects on Reissner-Nordstr\"{o}m-AdS Black Hole}
\author{\"{O}zg\"{u}r \"{O}kc\"{u}}
\email{ozgur.okcu@ogr.iu.edu.tr}
\author{Ekrem Ayd{\i}ner}

 \email{ekrem.aydiner@istanbul.edu.tr}
\affiliation{Department of
Physics, Faculty of Science, \.{I}stanbul University,
\.{I}stanbul, 34134, Turkey}%

\date{\today}

\begin{abstract}
In this paper, we study the effects of thermal fluctuations for Reissner-Nordstr\"{o}m-AdS (RN-AdS) black hole. We obtain the corrected thermodynamic quantities such as entropy, temperature, equation of state and heat capacities in the presence of thermal fluctuations. We also investigate the phase transition of RN-AdS black hole for thermal fluctuations. Finally we compute the critical exponents for small thermal fluctuations. We show that critical exponents are the same as critical exponents without thermal fluctuations.
\end{abstract}

\maketitle


\section{Introduction}
\label{intro}
Black holes behave as thermodynamic objects with a Bekenstein-Hawking entropy and a Hawking temperature \cite{Beken1972,Beken1973,Bard1973,Hawk1974,Beken1974,Hawk1975,Hawk1983}, and set deep connection between the laws of classical general relativity, thermodynamics, and quantum mechanics due to the presence of
horizons \cite{Beken1972,Beken1973,Bard1973,Hawk1974,Beken1974,Hawk1975,Hawk1983,Town1997,Wald2001,Page2005}. It is suggested that black holes have very massive entropy in the universe, i.e., their entropies are greater than any other object of the same volume \cite{Hawk1974,Hawk1975}. This  entropy is defined by $S = A/4$. However, it is found that quantum fluctuations on the planck scale may change the entropy of the black holes with time. These quantum fluctuations become important when the black hole shrinks its size  and therefore, entropy of the black holes need to get corrected. In order to get corrections on the black hole entropy different approaches such as non-perturbative quantum general relativity, the Cardy formula, the exact partition function, matter fields in backgrounds of a black hole, Rademacher expansion of the partition function and string theoretical effects have been used \cite{Asht1991,Mann1998,Solo1998,Med1999,Carlip2000,Govin2001,Birm2001,Med2001,Lowe2010,Jing2000,Sen2011,Sen2013}. On the other hand, recently  the corrections to the black hole entropy can be obtained through  the the generalized uncertainty principle \cite{Ali2012,Gan2014,Faizal2015}. 

Quantum fluctuations  may lead to thermal fluctuations in the black hole thermodynamics. Based on this assumption, in Ref. \cite{Das2002} shown that thermal fluctuations can also contribute to entropy of the black hole. They suggested that the logarithmic corrections to Bekenstein-Hawking entropy can be interpreted as corrections due to the thermal fluctuations of the black hole around its equilibrium configuration. Recently, the effects of thermal fluctuations were studied for different kind of black objects such as charged AdS black hole \cite{Pour2015}, charged hairy BTZ black hole \cite{Sade2014}, black Saturn \cite{Faizal2015a,Pour2016}, modified Hayward black hole \cite{Pour2016a}, dyonic charged AdS black hole \cite{Sade2016} and Kerr-AdS black hole \cite{Pour2016b}.

It is shown that the RN-AdS black holes have a van der Waals like first order small-large black hole phase transition ending in a critical point \cite{Cham1999,Cham1999a}. Moreover the phase transition in RN-AdS black hole is remarkable coincidence with van der Waals liquid-gas phase transition, when the cosmological constant and its conjugate quantities are considered as thermodynamical variables \cite{Kub2012}. The cosmological constant corresponds to pressure
\begin{equation}
	P=-\frac{\Lambda}{8 \pi}=\frac{3}{8 \pi}\frac{1}{l^{2}}
	\label{press}
\end{equation}
and its conjugate quantity corresponds to thermodynamic volume, $V=\frac{\partial M}{\partial P}$. For the RN-AdS black hole, it is given by
\begin{equation}
	V=\frac{4}{3}\pi r_{h}^{3}\, .
	\label{tV}
\end{equation}
Variable cosmological constant notion has led to a rich structure of phase transitions and critical phenomena. Indeed, in the recent years, thermodynamics and phase transition properties of the black holes in AdS space have been discussed in many studies \cite{Kub2012,Niu2012,Tsai2012,Bane2011,Bane2012,Cald2000,Kast2009,Dolan2011,Belhaj2013,Spa2013,Cai2013,Moa2013,Dutta2013,Alta2014,Dolan2011a,okcu2017,John2014,Maj2016}.

In this article, we investigate  the thermodynamics and phase transition of RN-AdS black hole with thermal fluctuations following the method in the Refs. \cite{Das2002,Kub2012}. As mentioned before, authors studied effects of thermal fluctuations for RN-AdS black holes \cite{Pour2015} and  our results differ from Ref. \cite{Pour2015} which follows different approaches and assumptions. 

The paper is organized as follows. In section 2, we briefly review the logarithmic correction to black hole entropy due to thermal fluctuations \cite{Das2002}. In section 3, we obtain corrected thermodynamic quantities and investigate phase transition of the RN-AdS black hole in the presence of thermal fluctuations. We also compute critical exponents of the phase transition with small thermal fluctuations. Finally, we discuss our results in section 4. (We use the units $G_{N} = \hbar = k_{B} = c = 1$.)

\section{Logaritmic Corrections to Black Hole Entropy}
\label{LC}
In this section, we will briefly review the thermal fluctuations correction to black hole entropy \cite{Das2002,Pour2015}. First, we begin to consider partition function of a canonical ensemble
\begin{equation}
	Z(\beta)=\intop_{0}^{\infty}\rho(E)e^{-\beta}dE\, ,
	\label{cE}
\end{equation}
where $\beta$ denotes the inverse of the temperature, $\beta=\frac{1}{T}$. One can obtain density of states from Eq. (\ref{cE}),
\begin{equation}
	\rho(E)=\frac{1}{2\pi i}\intop_{c-i\infty}^{c+i\infty}Z(\beta)e^{\beta E}d\beta=\frac{1}{2\pi i}\intop_{c-i\infty}^{c+i\infty}e^{S(\beta)}d\beta\, ,
	\label{DoS}
\end{equation}
where 
\begin{equation}
	S(\beta)=lnZ(\beta)+\beta E\, .
	\label{ES}
\end{equation}
Expanding $S(\beta)$ around the equilibrium temperature $\beta_{0}$ , one can obtain
\begin{equation}
	S=S_{0}+\frac{1}{2}(\beta-\beta_{0})^{2}\left(\frac{\partial^{2}S(\beta)}{\partial\beta^{2}}\right)_{\beta=\beta_{0}}+...\, ,
	\label{SE}
\end{equation}
where $S_{0}=S(\beta_{0})$. From Eqs. (\ref{DoS}) and (\ref{SE}), density of state is given by
\begin{equation}
	\rho(E)=\frac{e^{S_{0}}}{2\pi i}\intop_{c-i\infty}^{c+i\infty}e^{\frac{1}{2}(\beta-\beta_{0})^{2}S_{0}^{''}}d\beta\,
	\label{NDoS}
\end{equation}
where $S_{0}^{''}=\left(\frac{\partial^{2}S(\beta)}{\partial\beta^{2}}\right)_{\beta_{0}}$. Substituting $c=\beta_{0}$ and $(\beta-\beta_{0})=iz$ in Eq. (\ref{NDoS}), density of state is given by
\begin{equation}
	\rho(E)=\frac{e^{S_{0}}}{\sqrt{2\pi S_{0}^{''}}}
	\label{NDoS2}
\end{equation}
and we can write
\begin{equation}
	S=S_{0}-\frac{1}{2}lnS_{0}^{''}+... \, .
	\label{ent2}
\end{equation}
This formula is valid for all thermodynamic system considered as canonical ensemble and it is also valid for black holes. $S_{0}$ denotes the black hole entropy and thus we can replace $T \rightarrow T_{H}$. Now, we can determine $S_{0}^{''}$. We consider the entropy, which is suggested in Ref. \cite{Das2002} and based on Ref. \cite{Carlip2000}, has the form,  
\begin{equation}
	S=x\beta^{m}+y\beta^{-n}\, ,
	\label{AE}
\end{equation}
where $x, y, m, n>0$. This entropy has an extremum at
\begin{equation}
	\beta_{0}=\left(\frac{ny}{mx}\right)^{\frac{1}{m+n}}\, .
	\label{ext}
\end{equation}
From Eq.(\ref{AE}) and Eq.(\ref{ext}), one can obtain 
\begin{equation}
	S(\beta_{0})=x\left(\frac{ny}{mx}\right)^{\frac{m}{m+n}}+y\left(\frac{mx}{ny}\right)^{\frac{n}{m+n}}\, , 
	\label{eT}
\end{equation}
\begin{equation}
	S^{''}(\beta_{0})=m(m-1)x\left(\frac{ny}{mx}\right)^{\frac{m-2}{m+n}}+n(n+1)y\left(\frac{mx}{ny}\right)^{\frac{n+2}{m+n}}
	\label{eT2}\, .
\end{equation}
Using Eqs. (\ref{eT}) and (\ref{eT2}) in Eq.(\ref{SE}), one can give the entropy as
\begin{equation}
	S(\beta)=A(x^{n}y^{m})^{\frac{1}{m+n}}+\frac{1}{2}B(x^{n+2}y^{m-2})^{\frac{1}{m+n}}(\beta-\beta_{0})^{2}+... \, 
	\label{expE4}
\end{equation}
where
\begin{equation}
	A=\left(\frac{n}{m}\right)^{\frac{m}{m+n}}+\left(\frac{m}{n}\right)^{\frac{n}{m+n}}, \quad B=(m+n)m^{\frac{n+2}{m+n}}n^{\frac{m-2}{m+n}} 
\end{equation}
are the constants. Comparing with Eq.(\ref{SE}), we can find
\begin{equation}
	S(\beta_{0})=A(x^{n}y^{m})^{\frac{1}{m+n}}, \quad S^{''}(\beta_{0})=B(x^{n+2}y^{m-2})^{\frac{1}{m+n}}\, .
	\label{expE5}
\end{equation}
We can obtain $x$ and $y$ in terms of $S_{0}$ and $S^{''}_{0}$ from Eq.(\ref{expE5})
\begin{equation}
	x=\frac{A^{\frac{m-2}{2}}}{B^{\frac{m}{2}}}(S_{0}^{''})^{\frac{m}{2}}S_{0}^{-\frac{m-2}{2}}, \quad y=\frac{A^{-\frac{n+2}{2}}}{B^{-\frac{n}{2}}}S_{0}^{\frac{n+2}{2}}(S_{0}^{''})^{-\frac{n}{2}}\, .
\end{equation}
Substituting $x$ and $y$ in Eq.(\ref{ext}), we can obtain $\beta_{0}$ in terms of $S_{0}$, $S_{0}^{''}$
\begin{equation}
	\beta_{0}=\left(\frac{n}{m}\right)^{\frac{1}{m+n}}\sqrt{\frac{B}{A}\frac{S_{0}}{S_{0}^{''}}} \,
\end{equation}
and $S_{0}^{''}$ is given by
\begin{equation}
	S_{0}^{''}=\left[\left(\frac{B}{A}\right)\left(\frac{n}{m}\right)^{\frac{2}{m+n}}\right]S_{0}\beta_{0}^{-2}
\end{equation}
The factor in the square brackets can be negligible, so we can write
\begin{equation}
	S_{0}^{''}=S_{0}\beta_{0}^{-2}\, .
	\label{SDoE}
\end{equation}
Substituting Eq. (\ref{SDoE}) in Eq. (\ref{ent2}), we can obtain
\begin{equation}
	S=S_{0}-\frac{1}{2}lnS_{0}T_{H}^{2}\, .
	\label{TFS}
\end{equation}
This is the generic correction formula for black hole entropy. We will discuss, in the next section, the corrections to thermodynamic quantities due to thermal fluctuations.

\section{Reissner-Nordst\"{o}rm-AdS Black Hole}
\label{RNAdS}

RN-AdS black hole in four dimensional space is given by the metric
\begin{equation}
	ds^{2}=-f(r)dt^{2}+f^{-1}(r)dr^{2}+r^{2}d\theta^{2}+r^{2}\sin^{2}(\theta)d\phi^{2}\, ,
	\label{metric}
\end{equation}
with
\begin{equation}
	f(r)=1-\frac{2M}{r}+\frac{Q^{2}}{r^{2}}+\frac{r^{2}}{l^{2}}
	\label{metricFunc}
\end{equation}
where $l$, $M$ and $Q$ are the AdS radius, mass and charge of the black hole, respectively. Black hole event horizon $r_{h}$ is given by as a largest root of $f(r_{h})=0$. The mass of black hole can be obtained in terms of $r_{h}$, $Q$ and $l$ from Eq. (\ref{metricFunc}),
\begin{equation}
	M=\frac{r_{h}}{2}\left(1+\frac{Q^{2}}{r_{h}^{2}}+\frac{r_{h}^{2}}{l^{2}}\right)\, .
	\label{mass}
\end{equation}
Hawking temperature is given by
\begin{equation}
	T_{H}=\left(\frac{f'(r_{h})}{4\pi}\right)_{r=r_{h}}=\frac{1}{4\pi r_{h}}\left(1+\frac{3r_{h}^{2}}{l^{2}}-\frac{Q^{2}}{r_{h}^{2}}\right)
	\label{HT}
\end{equation}
and expression for the entropy can be written
\begin{equation}
	S_{0}=\frac{A}{4}=\pi r_{h}^{2}, \quad A=4\pi r_{h}^{2}\, .
	\label{Ent}
\end{equation}
In Ref. \cite{Pour2015} to track corrections coming from the thermal fluctuations, adding $a$ parameter into Eq. (\ref{TFS}), entropy is generalized as
\begin{equation}
	S=S_{0}-\frac{a}{2}lnS_{0}T_{H}^{2}\, .
	\label{correctEnt}
\end{equation}
In the case of $a = 1$, it is assumed that thermal fluctuations contribution is maximum, however, by setting $a=0$, we will recover entropy expression in Eq. (\ref{Ent}) without any corrections. Indeed, if we can use Eqs. (\ref{HT}) and (\ref{Ent}) into Eq. (\ref{correctEnt}),
we can obtain the following expression,
\begin{equation}
	S=\pi r_{h}^{2}-a\ln\left(\frac{l^{2}(r_{h}^{2}-Q^{2})+3r_{h}^{4}}{4\sqrt{\pi}l^{2}r_{h}^{2}}\right)
	\label{CorrectEnt}
\end{equation}
which refers to new entropy definition of black hole due to thermal fluctuations. For $a=0$, entropy in Eq. (\ref{Ent}) is recovered. Corrected temperature can be obtained from $T=\frac{\partial M}{\partial S}$
\begin{equation}
	T=\frac{(3r_{h}^{4}+l^{2}(r_{h}^{2}-Q^{2}))^{2}}{4l^{2}r_{h}(3\pi r_{h}^{6}+(\pi l^{2}-3a)r_{h}^{4}-\pi l^{2}Q^{2}r_{h}^{2}-l^{2}Q^{2}a)}\, .
	\label{CorrectT}
\end{equation}
Specific heat can be obtained from $C=T\frac{\partial S}{\partial T}$ and thus specific heats at constant volume and pressure can be given by
\begin{equation}
	C_{V}=a\frac{3\pi r_{h}^{6}+(\pi l^{2}-3a)r_{h}^{4}-\pi l^{2}Q^{2}r_{h}^{2}-al^{2}Q^{2}}{-3\pi r_{h}^{6}+(3a-\pi l^{2})r_{h}^{4}+l^{2}(\pi Q^{2}-a)r_{h}^{2}+3al^{2}Q^{2}}
	\label{CV}	
\end{equation}
and
\begin{equation}
	C_{p}=2\frac{9\pi^{2}r_{h}^{12}+6\pi^{2}l^{2}r_{h}^{10}+\pi^{2}l^{2}(l^{2}-6Q^{2})r_{h}^{8}-2\pi^{2}l^{4}Q^{2}r_{h}^{6}+\pi^{2}l^{4}Q^{4}r_{h}^{4}+\chi_{1}(a)}{9\pi r_{h}^{10}+\pi l^{2}(6Q^{2}-l^{2})r_{h}^{6}+4\pi l^{4}Q^{2}r_{h}^{4}-3\pi l^{4}Q^{4}r_{h}^{2}+\chi_{2}(a)}
\end{equation}
where $\chi_{1}(a)=-18a\pi r_{h}^{10}+3a(3a-2\pi l^{2})r_{h}^{8}+2a l^{2}Q^{2}(3a-\pi l^{2}Q^{2})r_{h}^{4}+2a\pi l^{4}Q^{4}r_{h}^{2}+a^{2}l^{4}Q^{4}$ and $\chi_{2}(a)=3a(-9r_{h}^{8}+l^{2}r_{h}^{6}-12Q^{2}l^{2}r_{h}^{4}-l^{4}Q^{2}r_{h}^{2})$. 

Local thermodynamic stability of canonical ensemble is given by $C_{P}>0$ \cite{Alta2014}. If $C_{P}<0$, black hole is thermodynamically unstable.
\begin{figure}[b]
	\centerline{\includegraphics[width=10cm]{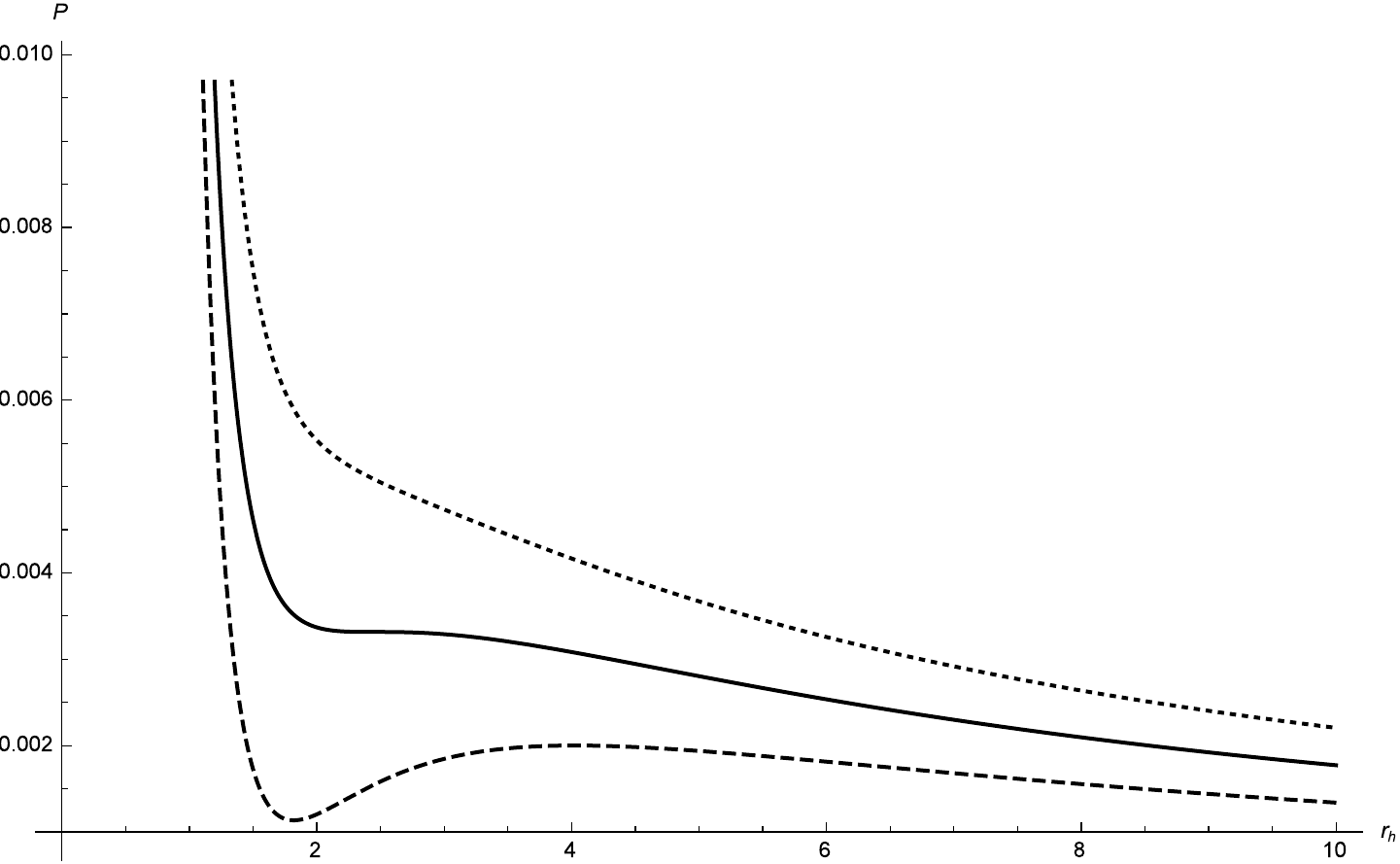}}
	\caption{The $P-r_{h}$ diagram. The temperatures of the isotherms decreases from top to bottom and correspond to $1.2T_{c}$, $T_{c}$, $0.8T_{c}$. We fix $a=10^{-9}$ and $Q=1$. \label{fi1}}
\end{figure}
\begin{figure}[b]
	\centerline{\includegraphics[width=10cm]{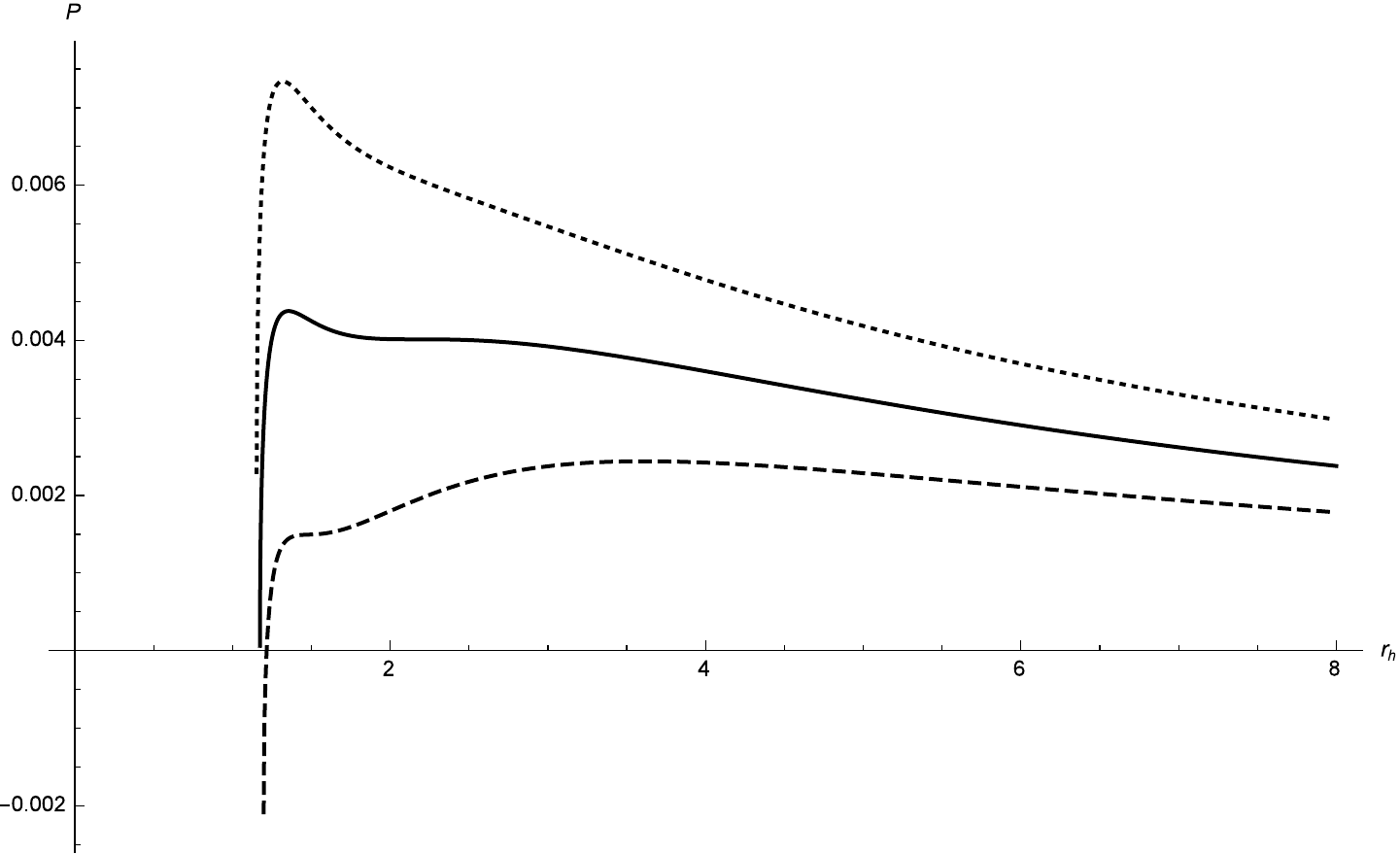}}
	\caption{The $P-r_{h}$ diagram. The temperatures of the isotherms decreases from top to bottom and correspond to $1.2T_{c}$, $T_{c}$, $0.8T_{c}$. We fix $a=1$ and $Q=1$. \label{fi2}}
\end{figure}
\begin{figure}[b]
	\centerline{\includegraphics[width=10cm]{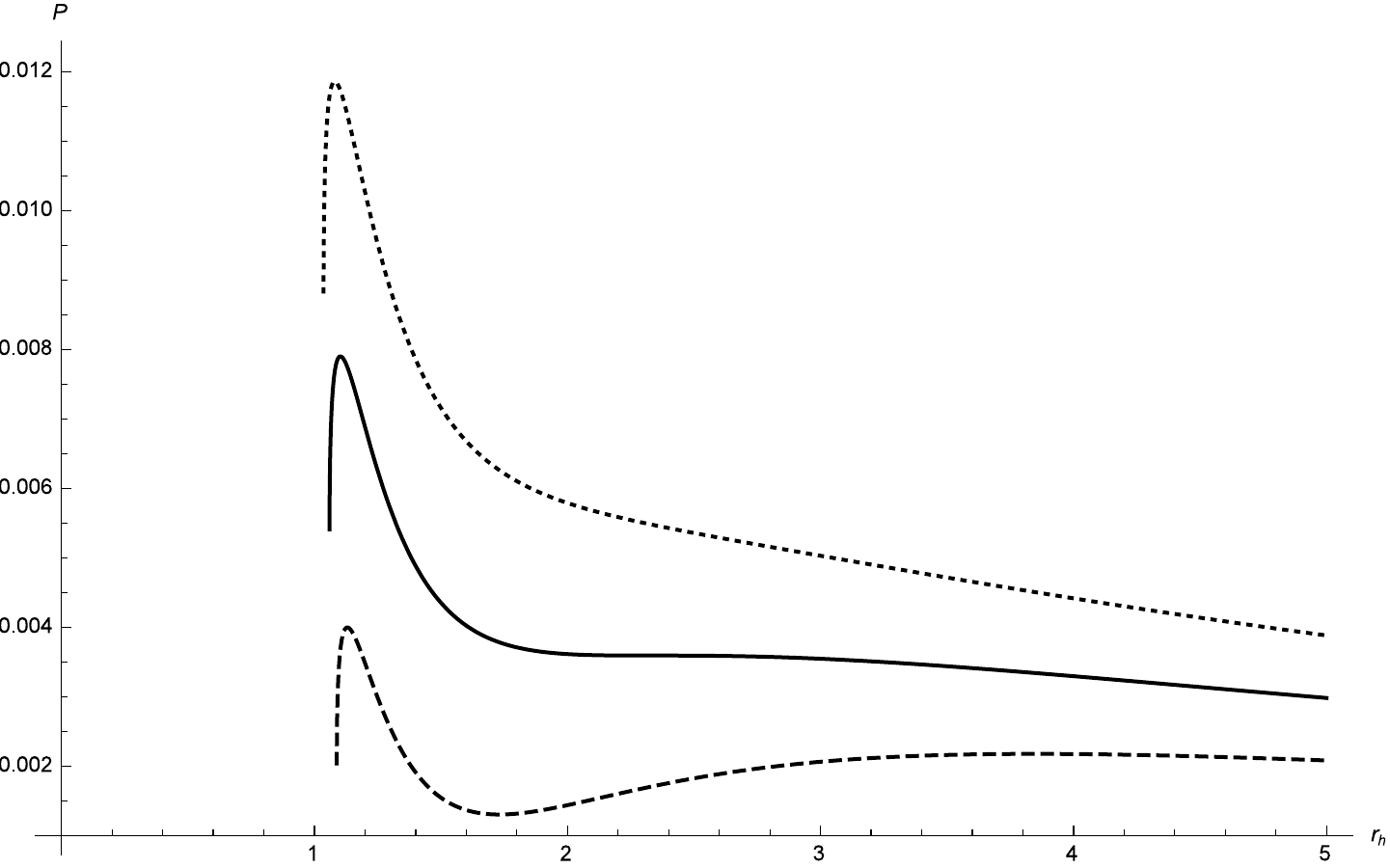}}
	\caption{The $P-r_{h}$ diagram. The temperatures of the isotherms decreases from top to bottom and correspond to $1.2T_{c}$, $T_{c}$, $0.8T_{c}$. We fix $a=0.5$ and $Q=1$. \label{fi3}}
\end{figure}
\subsection{Equation of State and Phase Transition}

In this section, we analyze the $P-V$ criticality of RN-AdS black holes in the presence of thermal fluctuations. We use the critical point investigation method which is defined in Ref. \cite{Kub2012}. From Eqs. (\ref{press}) and (\ref{CorrectT}), one can obtain the equation of state
\begin{equation}
	P=\frac{Q^{2}}{8\pi r_{h}^{4}}-\frac{1}{8\pi r_{h}^{2}}+\frac{T}{4r_{h}}-\frac{aT}{4\pi r_{h}^{3}}\pm\frac{\sqrt{\pi^{2}T^{2}r_{h}^{14}-W(a)}}{4\pi r_{h}^{8}}
	\label{EOS1}
\end{equation}
where $W(a)=aTr_{h}^{9}(2\pi Tr_{h}^{3}-r_{h}^{2}-aTr_{h}+2Q^{2})$. The positive sign has been taken before the square root so we can obtain the equation of state in the $a\rightarrow 0$ limit. The critical points are obtained from
\begin{equation}
	\frac{\partial P}{\partial r_{h}}=0, \quad \frac{\partial^{2}P}{\partial r_{h}^{2}}=0\, .
\end{equation}
It is possible to obtain critical points for the small values of $a$. For small thermal fluctuations, equation of state can be expanded as
\begin{equation}
	P=\frac{Q^{2}}{8\pi r_{h}^{4}}-\frac{1}{8\pi r_{h}^{2}}+\frac{T}{2r_{h}}+a\left(\frac{1}{8\pi^{2}r_{h}^{4}}-\frac{T}{2\pi r_{h}^{3}}-\frac{Q^{2}}{4\pi^{2}r_{h}^{6}}\right)+\mathcal{O}(a^{2})\, .
	\label{smallTher}
\end{equation}
Specific volume $v$ should be identified with
\begin{equation}
	v=2r_{h}
	\label{SV}
\end{equation}
for the small thermal fluctuations. Now one can obtain the critical points for small thermal fluctuations
\begin{equation}
	v_{c}=2\sqrt{6}Q-\frac{\sqrt{6}}{2\pi Q}a+\mathcal{O}(a^{2})\, ,
	\label{vc}
\end{equation}
\begin{equation}
	T_{c}=\frac{\sqrt{6}}{18\pi Q}+\frac{\sqrt{6}}{72\pi^{2}Q^{3}}a+\mathcal{O}(a^{2})\, ,
	\label{Tc}
\end{equation}
\begin{equation}
	P_{c}=\frac{1}{96\pi Q^{2}}+\frac{a}{216\pi^{2}Q^{4}}+\mathcal{O}(a^{2})\, .
	\label{Pc}
\end{equation}
Universal number can be calculated from Eqs. (\ref{vc}), (\ref{Tc}) and (\ref{Pc})
\begin{equation}
	\frac{P_{c}T_{c}}{v_{c}}=\frac{3}{8}-\frac{a}{48\pi Q^{2}}+\mathcal{O}(a^{2})\, .
	\label{universal}
\end{equation}
As it can be seen from Eqs. (\ref{vc}), (\ref{Tc}), (\ref{Pc}) and (\ref{universal}) for $a = 0$, the results in Ref. \cite{Kub2012} are obtained i.e, when $a=0$, the critical points are $v_{c}=2\sqrt{6}Q$, $T_{c}=\frac{\sqrt{6}}{18\pi Q}$, $P_{c}=\frac{1}{96\pi Q^2}$
and  $\frac{P_{c}T_{c}}{v_{c}}=\frac{3}{8}$. Critical points change due to thermal fluctuations. For example critical specific volume decreases while the critical temperature and pressure increase. Universal number of phase transition depends on $Q$ and $a$.

In order to investigate the phase transition behavior of RN-AdS black hole, $P-r_{h}$ diagram has been plotted in Fig. \ref{fi1} for small values of thermal fluctuations. As it can be seen from the figure that under the critical temperature $P-r_{h}$ diagram has characteristic behavior of van der Waals phase transition. It is also important to investigate the phase transition behavior for large values of thermal fluctuations since effects of thermal fluctuations are important for small size. When thermal fluctuations are increased, the phase transition deviates from van der Waals behavior. In Figs. \ref{fi2} and \ref{fi3}, we have plotted $P-r_{h}$ diagrams for thermal fluctuations. It can easily be  seen from figures that thermal fluctuations affect the phase transition. Moreover, in contrast to Fig. \ref{fi1}, small black hole branch is unstable and pressure is imaginary under a certain event horizon radius. It may be interpreted as black holes do not exist under a certain value of event horizon. This interpretation is also useful to avoid the singularities and negative regions of entropy and temperature. For example, $1.2T_{c}$ isotherm in Fig. \ref{fi3} is imaginary for $r_{h}<1.03573$. Moreover, event horizons of entropy and temperature that become singular, are less than $r_{h}<1.03573$. It may imply that black hole never approaches ill-defined regions. 

\subsection{Critical Exponents}

The thermodynamic behavior of a system near the phase transition point has been classified by using critical exponents. It is supposed that the critical exponents to be universal and are the independent of the details of the interaction. On the other hand, different physical systems may share the same critical exponents, which indicates that they are in the same universal class. In this section, we discuss critical exponents of the phase transition at near the critical point for the RN-AdS black hole. The critical exponents are given as follows:

The exponent $\alpha$ determines the behavior of specific heat at constant volume
\begin{equation}
	C_{V}=T\left(\frac{\partial S}{\partial T}\right) \propto |t|^{-\alpha}\, .
	\label{alpha}
\end{equation}
The exponent $\beta$ determines the behavior of the order parameter on the given isotherm
\begin{equation}
	\eta=V_{l}-V_{s} \propto |t|^{\beta}\, .
	\label{beta}
\end{equation}
The exponent $\gamma$ determines the behavior of the isothermal compressibility $\kappa_{T}$
\begin{equation}
	\kappa_{T}=-\frac{1}{V}\left(\frac{\partial V}{\partial P}\right) \propto |t|^{-\gamma}\, .
	\label{gamma}
\end{equation}
The exponent $\delta$ determines the following behavior on critical isotherm $T=T_{c}$
\begin{equation}
	|P-P_{c}| \propto |V-V_{c}|^{\delta}\, .
	\label{delta}
\end{equation}
Recently, a more general method is suggested to obtain critical exponents in Ref. \cite{Maj2016}. However, in our case, it is convenient to use method in Ref. \cite{Kub2012}.  Now following method in Ref. \cite{Kub2012}, we can obtain critical exponents of phase transition for RN-AdS black hole. We cannot obtain the exact critical point due to the complexity of equation of state in Eq. (\ref{EOS1}) so we cannot obtain the law of corresponding states from Eq. (\ref{EOS1}) the critical exponent may differ from Ref. \cite{Kub2012}. On the other hand, we can compute the critical exponents for small thermal fluctuations. In the presence of small thermal fluctuations, we can obtain specific heat at constant volume, $C_{V}=-a+\mathcal{O}(a^{2})$ and this is independent of $T$ so exponent $\alpha=0$. Defining
\begin{equation}
	p=\frac{P}{P_{c}},\quad \tau=\frac{T}{T_{c}},\quad \nu=\frac{V}{V_{c}}\, ,
	\label{cS}
\end{equation}
we can obtain the so called law of corresponding states with small thermal fluctuations. Expanding around the critical point
\begin{equation}
	t=\tau-1,\quad \omega=\nu-1
	\label{ND}
\end{equation}
we find equation of state
\begin{equation}
	p=1+\left(\frac{8}{3}-\frac{8a}{27\pi Q^{2}}\right)t+\left(-\frac{8}{9}+\frac{32a}{81\pi Q^{2}}\right)t\omega+\left(-\frac{4}{81}+\frac{16a}{729\pi Q^{2}}\right)\omega^{3}+\mathcal{O}(t\omega^{2},\omega^{4})\, .
	\label{LCS}
\end{equation}
Differentiating Eq.(\ref{LCS}) for a fixed $t < 0$ we obtain
\begin{equation}
	dP=-\frac{4}{27}P_{c}\left[\left(6-\frac{8a}{3\pi Q^{2}}\right)t+\left(1-\frac{4a}{9\pi Q^{2}}\right)\omega^{2}\right]d\omega\, .
	\label{DLCS}
\end{equation}
Employing Maxwell’s equal area law, one can get the following two equations:
\begin{eqnarray}\label{a40}
	&p=1+\left(\frac{8}{3}-\frac{8a}{27\pi Q^{2}}\right)t+(-\frac{8}{9}+\frac{32a}{81\pi Q^{2}})t\omega_{l}+(-\frac{4}{81}+\frac{16a}{729\pi Q^{2}})\omega_{l}^{3}\nonumber\\ &=1+\left(\frac{8}{3}-\frac{8a}{27\pi Q^{2}}\right)t+(-\frac{8}{9}+\frac{32a}{81\pi Q^{2}})t\omega_{s}+(-\frac{4}{81}+\frac{16a}{729\pi Q^{2}})\omega_{s}^{3}\, ,\nonumber \\
	&0=\oint\omega dP=\intop_{\omega_{l}}^{\omega_{s}}\omega\left[\left(6-\frac{8a}{3\pi Q^{2}}\right)t+\left(1-\frac{4a}{9\pi Q^{2}}\right)\omega^{2}\right]d\omega\, . 
	\label{solu}
\end{eqnarray}
where $\omega_{s}$ and $\omega_{l}$ denotes volume of small and large black holes. From these equation, one can obtain
\begin{equation}
	\omega_{l}=-\omega_{s}=3\sqrt{-2t}\, .
\end{equation}
So we find the exponent $\beta$
\begin{equation}
	\eta=V_{c}(\omega_{l}-\omega_{s})=2V_{c}\sqrt{-2t}\quad \Rightarrow \quad \beta=\frac{1}{2}\, .
	\label{betaS}
\end{equation}
We compute the isothermal compressibility exponent $\gamma$
\begin{equation}
	\kappa_{T}=-\frac{1}{V}\left(\frac{\partial V}{\partial P}\right)_{T}\propto\frac{81\pi Q^{2}}{8P_{c}(9\pi Q^{2}-4a)t} \quad \Rightarrow \quad \gamma=1\, .
	\label{gammaS}
\end{equation}
Finally we obtain the exponent $\delta$ for $t=0$
\begin{equation}
	p-1=\left(-\frac{4}{81}+\frac{16a}{729\pi Q^{2}}\right)\omega^{3} \quad \Rightarrow \quad \delta=3\, .
\end{equation}
We computed the critical exponents for small thermal fluctuations. All critical exponents that were computed in the presence of small thermal
fluctuations coincide critical exponents of van der Waals fluid and RN-AdS black hole without thermal fluctuations \cite{Kub2012}.

\section{Conclusion}
\label{result}
In this paper, we studied effects of thermal fluctuations for RN-AdS black hole. We obtained some corrected thermodynamic quantities and investigated $P-V$ criticality in the presence of thermal fluctuations. We obtained critical points for small thermal fluctuations and numerically studied phase transition for large values of thermal fluctuations. Thermal fluctuations effects are remarkable for small size black hole and also affect the phase transition. Black hole may not exist under a certain value of event horizon. This minimum event horizon condition provides us to avoid ill-defined regions of temperature and entropy. Finally we obtained the critical exponent for small thermal fluctuations. We showed that critical exponents are the same as critical exponent without thermal fluctuations.

\section{Conflict of Interest}
The authors declare that there is no conflict of interest regarding the publication of this paper.
\

\

\begin{acknowledgements}
This work was supported by Scientific Research Projects Coordination Unit of Istanbul University. Project numbers is FYL-2016-2061.
\end{acknowledgements}


\end{document}